\documentclass[doublecol]{epl2} 
\usepackage{amsmath}

\def\kk{\textbf{\textit{k}}}

\def\rr{\textbf{\textit{r}}}

\def\SS{\textbf{\textit{S}}}

\title{Thermal relaxation of magnons and phonons near resonance points in magnetic insulators}
\shorttitle{Thermal relaxation of magnons and phonons near resonance points} 

\author{Q. Xi\inst{1} \and B. Liu\inst{1} \and Z. Shi\inst{2}\thanks{E-mail: \email{shizhong@tongji.edu.cn}} \and T. Nakayama\inst{1,3} \and J. Zhou\inst{1}\thanks{E-mail: \email{zhoujunzhou@tongji.edu.cn}} \and B. Li\inst{4}}
\shortauthor{Q. Xi \etal}

\institute{                    
  \inst{1} Center for Phononics and Thermal Energy Science, China-EU Joint Lab for Nanophononics,
School of Physics Science and Engineering, Tongji University, 
Shanghai 200092, China\\
  \inst{2} Shanghai Key Laboratory of Special Artificial Microstructure Materials and Technology,
School of Physics Science and Engineering, Tongji University,
Shanghai 200092, China\\
  \inst{3} Hokkaido University, Sapporo, Hokkaido 060-0826, Japan\\
  \inst{4} Department of Mechanical Engineering, University of Colorado at Boulder, CO 80309, USA                            	
}

\abstract{We theoretically investigate the energy relaxation rate of
magnons and phonons near the resonance points to clarify the
underlying mechanism of heat transport in ferromagnetic materials.
We find that the simple two-temperature model is valid for the
one-phonon/one-magnon process, as the rate of energy exchange between
magnons and phonons is proportional to the temperature difference
between them, and it is independent of temperature in the high
temperature limit.
We found that the magnon-phonon relaxation time due to the
one-phonon/one-magnon interaction could be reduced to 1.48 $\mu s$ at
the resonance point by applying an external magnetic field. 
It means that the resonance effect plays a significant role in
enhancing the total magnon-phonon energy exchange rate, apart from the
higher order interaction processes.
}

\begin{document}

\maketitle

The heat transport in insulators is generally dominated by
phonons which are the quanta of lattice vibrations. In magnetic
insulators, the spin waves (magnons) could
also act as heat carriers\,\cite{Hess2003, Jin2003}, as observed for
yttrium iron garnet (YIG)\,\cite{Boona2014},
Nd$_2$CuO$_4$\,\cite{Li2005}, RbMnF$_3$\,\cite{Gustafson1973},  and
MnF$_2$\,\cite{Sanders1977}.
The notable contribution of magnons to thermal conductivity has also
been discovered in the spin ladder compound
(Sr,Ca,La)$_{14}$Cu$_{24}$O$_{41}$\,\cite{Hess2003,Hohensee2014,Sologubenko2000}.
In addition, the characteristics of heat transport due to magnons have
been employed to probe spin excitations in InGaAs quantum dot
system\,\cite{Lorenz2002}.
In recent years, the heat current due to spin excitations has also
stimulated the field of spin caloritronics\,\cite{Bauer2012}, 
resulting in the recent discovery of spin Seebeck
effect\,\cite{Xiao2010}, and spin Peltier effect\,\cite{Flipse2012}.

Intuitively, the total thermal conductivity ($\kappa_{T}$) of magnetic insulators
could be evaluated by a simple sum of the
lattice thermal conductivity ($\kappa_p$) and magnon thermal
conductivity ($\kappa_m$)
contributions: $\kappa_{T}=\kappa_p+\kappa_m$. However, magnons will be
scattered by phonons and vice versa, thus the interaction between
phonons and magnons becomes relevant in determining the total thermal
conductivity of magnetic materials. An effective strategy to evaluate
the heat transport through both phonons and magnons, together with
their interaction, is the two-temperature
model\,\cite{Agrawal2013,An2016}. It was first proposed by Sanders and
Walton for the coupled magnon-phonon mode diffusion in the ferrimagnet
YIG and the antiferromagnet MnF$_2$\,\cite{Sanders1977}. 
In their paper, phonons and magnons are assumed to be
excited to their equilibrium states with different effective
temperatures, with the local energy exchange rate between these two
carriers proportional to their temperature difference.   
They found that thermal conductivity is not determined only by
$\kappa_m$ and $\kappa_p$ but also by the magnon-phonon
relaxation time $\tau_{mp}$\cite{Sanders1977}.
Recently, Chen et al. has generalized a two-temperature model
including the effect of the concurrent magnetization flow by assuming
a constant magnon-phonon relaxation time \,\cite{Liao2014}.
However, there has been no theoretical work to check the validity of
the introduction of $\tau_{mp}$.
In addition, crucial problems such as the strength of the
magnetoelastic coupling, the exact relaxation
time\,\cite{Roschewsky2014, Schreier2013}, and their magnetic field dependence
have not been fully understood yet. 

Taking YIG as a prototype for studying the magnon properties in ferromagnetism\,\cite{Princep2017,Man2017}.
Sato\,\cite{Sato1955} suggested that the thermal
conductivity due to magnon proportional to $T^2$ could be greater than
that of phonons to $T^3$ below 1K by assuming that the mean free paths
of phonons and magnons are limited by the boundary condition and
comparable in magnitude.
Later on, Daugless\,\cite{Douglass1963} experimentally observed a large
decrease of thermal conductivity in YIG when applying a magnetic field
of 2T at 0.5 K. To explore the field and temperature dependence of thermal transport
properties, many theoretical and experimental researches have been
carried out. Those have revealed that the magnetoelastic coupling between phonons
and magnons plays a significant role for governing the total thermal
conductivity\,\cite{An2016,Bhandari1966,Walton1973}.
The strength of the coupling reaches a peak at the resonance
condition, where the one-phonon/one-magnon process with the phonon
and magnon of same frequency and wave vector becomes
relevant\,\cite{Kittel1958,Weber1968}. 

In this Letter, we focus our attention on the resonance behavior of the
one-phonon/one-magnon interaction. We derive a simple formula for the
rate of energy exchange between magnons and phonons in analogy with
the two-temperature model for electron-phonon
system\,\cite{Allen1987}.
We present the physical conditions to set up the two-temperature model
based on the one-phonon/one-magnon interaction process. The field
dependence of the temperature relaxation rate and the relaxation time
corresponding to the one-phonon/one-magnon interaction is also
obtained.

The magnetic Hamiltonian of YIG consists of dipolar, exchange
interactions between spins, and the Zeeman splitting due to the
external magnet field ($H$) along $z$ direction \cite{Flebus2017}:  
\begin{align}
\nonumber
H_{\rm{mag}}&=\frac{\mu_0\left(g\mu_{\rm{B}}\right)^2}{2}\sum_{i\neq j}\frac{|\rr_{ij}|^2\SS_i\cdot\SS_{j}-3\left(\rr_{ij}\cdot\SS_i\right)\left(\rr_{ij}\cdot\SS_j\right)}{|\rr_{ij}|^5} \\
&-J\sum_{i\neq j}\SS_{i}\cdot\SS_{j}-g\mu_BH\sum_{i}S_{i}^{z},
\label{equ1}
\end{align}
where $\mu_0$ is the vacuum permeability, $\mu_B$ is the Bohr
magneton, $g$ is the g factor, $J$ is the exchange
integral. The spin $\SS_{i}=\SS\left(\rr_i\right)$ locates on the site
$\rr_i$ with $S=|S_i|=a_0^3M_s/g\mu_B$, where $a_0$ is the unit cell
lattice constant, $M_s$ is the saturation magnetization density, and
$\rr_{ij}=\rr_i-\rr_j$. 
By employing the Holstein-Primakoff transformation, the quantized
Hamiltonian for spin excitations could be expressed
as\,\cite{Flebus2017}:
\begin{eqnarray}
H_{\rm{mag}}=\sum_{\kk}A_{\kk}a_{\kk}^{\dagger}a_{\kk}+\frac{1}{2}\left(B_{\kk}a_{-\kk}a_{\kk}+B_{\kk}^{\ast}a_{\kk}^{\dagger}a_{-\kk}^{\dagger}\right),
\label{equ2}
\end{eqnarray}
with
\begin{eqnarray}
\nonumber
\frac{A_{\kk}}{\hbar}&=&D\mathcal{F}_{\kk}+\gamma \mu_0 H+\frac{\gamma\mu_0M_s\sin^2\theta_{\kk}}{2},\\
\frac{B_{\kk}}{\hbar}&=&\frac{\gamma\mu_0M_s\sin^2\theta_{\kk}}{2}e^{-2i\phi_{\kk}},
\end{eqnarray}
where $a_{\kk}^{\dagger}(a_{\kk})$ is the  magnon creation
(annihilation) operators with wave vector $\kk$, $D=2SJa_0^2$ the
exchange stiffness, $\gamma=g\mu_B/\hbar$ the gyromagnetic ratio,
$\theta_{\kk}=\arccos\left(k_z/k\right)$ the polar angle between wave
vector $\kk$ and the magnetic field along $z$ direction, and $\phi$
the azimuthal angle of $\kk$ in $xy$ plane. In the long-wavelength
limit, the form factor $\mathcal{F}_{\kk}\approx
k^2$\,\cite{Flebus2017}. 
Eq.\,(\ref{equ2}) could be diagonalized by using the Bogoliubov transformation:
\begin{eqnarray}
\left[\begin{matrix}
a_{\kk}\\
a_{-\kk}^{\dagger}
\end{matrix}
\right]
=\left[
\begin{matrix}
u_{\kk} & -v_{\kk}\\
-v_{\kk}^{\ast} & u_{\kk}
\end{matrix}
\right]\left[
\begin{matrix}
\alpha_{\kk} \\
\alpha_{-\kk}^{\dagger}
\end{matrix}
\right],
\end{eqnarray}
with 
\begin{eqnarray}
u_{\kk}=\sqrt{\frac{A_{\kk}+\hbar\omega_{\kk}}{2\hbar\omega_{\kk}}},  v_{\kk}=\sqrt{\frac{A_{\kk}-\hbar\omega_{\kk}}{2\hbar\omega_{\kk}}}e^{2i\phi_{\kk}}.
\end{eqnarray}
The Hamiltonian is then simplified to 
\begin{eqnarray}
H_{\text{mag}}=\sum_{\kk}\hbar\omega_{\kk}\alpha_{\kk}^{\dagger}\alpha_{\kk},
\end{eqnarray}
and the dispersion relation for bulk magnons in the long-wavelength limit is
\begin{eqnarray}
\omega_{\kk}=\sqrt{Dk^2+\gamma\mu_0H}\sqrt{Dk^2+\gamma\mu_0\left(H+M_s\sin^2\theta_{\kk}\right)}
\end{eqnarray}
The Hamiltonian for one-phonon/one-magnon interaction process has also been derived by Flebus et al.\,\cite{Flebus2017}:

\begin{align}
\nonumber
H_{\text{int}}&=&\hbar n B_{\perp}\left(\frac{\gamma\hbar^2}{4M_s\bar{\rho}}\right)^{\frac{1}{2}}\sum_{\kk,\lambda}k\omega_{\kk \lambda}^{-\frac{1}{2}}e^{-i\phi}a_{\kk}\left(c_{-\kk \lambda}+c_{\kk\lambda}^{\dagger}\right) \\
&\times&\left(-i\delta_{\lambda 1}\cos2\theta_\kk+i\delta_{\lambda2}\cos\theta_\kk-\delta_{\lambda 3}\sin 2\theta_\kk\right)+H.c. ,
\label{equ8}
\end{align}
where $n=1/a_0^3$ is the number density of the magnetic particles in
the system, $B_{\perp}$ the magnetoelastic constants, $\bar{\rho}$ the
average mass density, $c_{\kk\lambda}^{\dagger}$($c_{\kk\lambda}$) the
phonon creation (annihilation) operators with wave vector
$\kk$. $\delta$ is the Kronecker delta, and $\lambda=1,2$
labels the transverse acoustic (TA) phonon, $\lambda=3$ labels the
longitudinal acoustic (LA) phonon. Under Debye approximation, the
phonon dispersion relation is $\omega_{\kk
  \lambda}=C_{\lambda}|\kk|$. Following the procedure of Bogoliubov
transformation, Eq. (\ref{equ8}) could be expressed in terms of the
magnon quasiparticles operators
$\alpha_{\kk}^{\dagger}$($\alpha_{\kk}$):
\begin{align}
\nonumber
H_{\text{int}}&=\hbar n B_{\perp}\left(\frac{\gamma\hbar^2}{4M_s\bar{\rho}}\right)^{\frac{1}{2}}\sum_{\kk,\lambda}k\omega_{\kk \lambda}^{-\frac{1}{2}}e^{-i\phi}\left(u_{\kk}\alpha_{\kk}-v_{\kk}\alpha^{\dagger}_{-\kk}\right) \\
&\times\left(c_{-\kk
  \lambda}+c_{\kk\lambda}^{\dagger}\right)\left(-i\delta_{\lambda
  1}\cos2\theta_\kk+i\delta_{\lambda2}\cos\theta_\kk-\delta_{\lambda
  3}\sin 2\theta_\kk\right)\nonumber \\
&+H.c. .
\label{equ9}
\end{align}
The decay rate of the magnon and phonon distribution functions $n_{\text{m}}(\kk)$ and $n_{\text{p}}(\kk\lambda)$ are:
\begin{widetext}
\begin{subequations}
\begin{align}
\frac{\partial n_{\text{m}}\left(\kk\right)}{\partial t}&=-\frac{2\pi}{\hbar^2}\sum_{\lambda}|M_{\kk, \kk \lambda}|^2\delta\left(\omega_{\text{m}}-\omega_{\text{p}}\right) \times \{n_{\text{m}}\left(\kk\right)\left[1+n_{\text{p}}\left(\kk,\lambda\right)\right] -\left[1+n_{\text{m}}\left(\kk\right)\right]n_{\text{p}}\left(\kk\lambda\right)\} 
\label{equ10a}
\end{align}	
\begin{align}
\frac{\partial n_{\text{p}}\left(\kk\lambda\right)}{\partial t}&=-\frac{2\pi}{\hbar^2}|M_{\kk, \kk \lambda}|^2\delta\left(\omega_{\text{m}}-\omega_{\text{p}}\right)\times \{n_{\text{p}}\left(\kk\lambda\right)\left[1+n_{\text{m}}\left(\kk\right)\right]-\left[1+n_{\text{p}}\left(\kk\lambda\right)\right]n_{\text{m}}\left(\kk\right)\} 
\label{equ10b}
\end{align}	
\end{subequations}
\end{widetext}
\begin{floatequation}
\mbox{\textit{see eq.~\eqref{equ10a}                                                                                                                                                                                                                                                                                                                                                                                                                                                                                                                                                                    and eq.~\eqref{equ10b}                                                                                                                                                                                                                                                                                                                                                                                                                                                                                                                                                                     }}
\end{floatequation}
where $|M_{\kk, \kk \lambda}|^2=\frac{\hbar^4 n^2 B_{\perp}^2\gamma}{4M_{\text{s}}\bar{\rho}}k^2\omega_{\kk\lambda}^{-1}\beta_{\lambda} 
$, $\beta_{1}=|u_{\kk}+v_{\kk}|^2 \cos^2 2\theta_{\kk}$,
$\beta_{2}=|u_{\kk}+v_{\kk}|^2 \cos^2 \theta_{\kk}$,
$\beta_{3}=|u_{\kk}-v_{\kk}|^2\sin^2 2\theta_{\kk}$.
In this model, we assumed that other collision processes such as
phonon-phonon interaction and magnon-magnon interaction are strong enough
to keep the local equilibrium, then the distribution functions $n_{\text{m}}(\kk)$ and
$n_{\text{p}}({\kk\lambda})$ can be replaced by the equilibrium ones
$\{\exp\left[\hbar\omega(\kk)/k_{B}T_{\text{m}}\right]-1\}^{-1}$ and
$\exp\left[\hbar\omega(\kk\lambda)/k_{B}T_{\text{p}}\right]-1\}^{-1}$
where magnon and phonon temperatures
are noted as $T_{\text{m}}$ and $T_{\text{p}}$, respectively.

The energy of magnons and phonons are
$E_{\text{m}}=\sum_{\kk}\hbar\omega_{\kk}n_{\text{m}}\left(\kk\right)$
and
$E_{\text{p}}=\sum_{\kk\lambda}\hbar\omega_{\kk\lambda}n_{\text{p}}\left(\kk\lambda\right)$,
thus the changing rate of energy becomes:
\begin{widetext}
\begin{eqnarray}
\frac{\partial E_{\text{m}}}{\partial t}&=&-\frac{\partial E_{\text{p}}}{\partial t}
=-\frac{2\pi}{\hbar^2}\sum\limits_{\kk\lambda} \hbar\omega_{\text{m}}(\kk)|M_{\kk, \kk \lambda}|^2\delta\left(\omega_{\text{m}}-\omega_{\text{p}}\right)\left[n_{\text{m}}\left(\kk\right)-n_{\text{p}}\left(\kk\lambda\right)\right]\nonumber\\.
\label{equ13}
\end{eqnarray} 
\end{widetext}
\begin{floatequation}
\mbox{\textit{see eq.~\eqref{equ13}                                                                                                                                                                                                                                                                                                                                                                                                                                                                                                                                                                     }}
\end{floatequation}

A Taylor expansion of $n_{\text{m}}(\omega)-n_{\text{p}}(\omega)$ in terms of $T_{\text{m}}-T_{\text{p}}$ is
\begin{widetext}
\begin{align}
\label{equ14}
\nonumber
n_{\text{m}}(\omega)-n_{\text{p}}(\omega)=\frac{z^2e^z}{\left(e^z-1\right)^2}\frac{k_B}{\hbar\omega}\left(T_{\text{m}}-T_{\text{p}}\right)+\frac{1}{3!}\left[\frac{3e^zz^4}{2(e^z-1)^2}-\frac{3e^{2z}z^5}{(e^z-1)^3}+\frac{3e^zz^5}{2(e^z-1)^2} \right. \\
\left. +\frac{3e^{3z}z^6}{2(e^z-1)^4}-\frac{3e^{2z}z^6}{2(e^z-1)^3}+\frac{e^{z}z^6}{4(e^z-1)^2}\right]\left(\frac{k_B}{\hbar\omega}\right)^3\left(T_{\text{m}}-T_{\text{p}}\right)^3+...,
\end{align}
\end{widetext}
\begin{floatequation}
\mbox{\textit{see eq.~\eqref{equ14}                                                                                                                                                                                                                                                                                                                                                                                                                                                                                                                                                                     }}
\end{floatequation}
where $z=\hbar\omega/k_BT$ and
$T=(T_{\text{m}}+T_{\text{p}})/2$. Since the first term in the right
hand side of Eq.\,(\ref{equ14}) is much larger than other higher order
terms, only the linear term is enough in the evaluation. Therefore,
the energy changing rate is linearly dependent on the temperature
difference, and the temperature changing rate becomes: 
\begin{subequations}
\begin{eqnarray}
\frac{\partial T_{\text{m}}}{\partial t}=g_{\text{mp}}(T)\left(T_{\text{p}}-T_{\text{m}}\right),
\end{eqnarray}
\begin{eqnarray}
\frac{\partial T_{\text{p}}}{\partial t}=g_{\text{pm}}(T)\left(T_{\text{m}}-T_{\text{p}}\right),
\end{eqnarray}
\end{subequations}
with 
\begin{subequations}
\begin{eqnarray}
g_{\text{mp}}(T)=\frac{\frac{2\pi k_B}{\hbar^2}\sum\limits_{\kk\lambda} |M_{\kk, \kk \lambda}|^2\delta\left(\omega_{\text{m}}-\omega_{\text{p}}\right)\frac{z^2e^z}{(e^z-1)^2}}{C_{\text{m}}(T)},
\end{eqnarray}
\begin{eqnarray}
g_{\text{pm}}(T)=\frac{\frac{2\pi k_B}{\hbar^2}\sum\limits_{\kk\lambda} |M_{\kk, \kk \lambda}|^2\delta\left(\omega_{\text{m}}-\omega_{\text{p}}\right)\frac{z^2e^z}{(e^z-1)^2}}{C_{\text{p}}(T)}.
\end{eqnarray}
\label{Equ16}
\end{subequations}
At the same time, the temperature difference between $T_{\text{m}}$ and $T_{\text{p}}$ will decay exponentially, that
\begin{eqnarray}
\frac{\partial}{\partial t}\Delta T=-\frac{\Delta T}{\tau_{\text{mp}}(T)},
\end{eqnarray}  
where $\Delta T=T_{\text{m}}-T_{\text{p}}$ and $\tau_{\text{mp}}(T)=\left[g_{\text{mp}}(T)+g_{\text{pm}}(T)\right]^{-1}$.
\begin{figure}[htb]
\includegraphics[width=\linewidth]{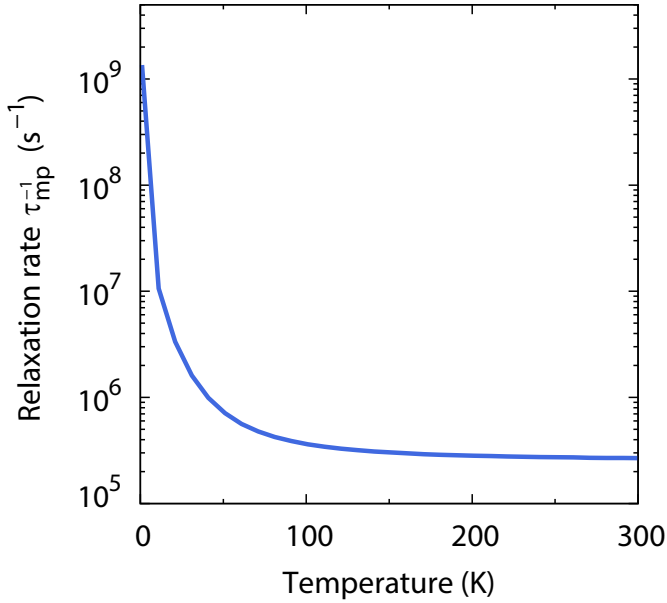}
\caption{(Color online) The magnon-phonon relaxation rate as a
  function of temperature when $\mu_0H=0$T.}
\label{fig_temp}
\end{figure}

\begin{figure}[htb]
\includegraphics[width=\linewidth]{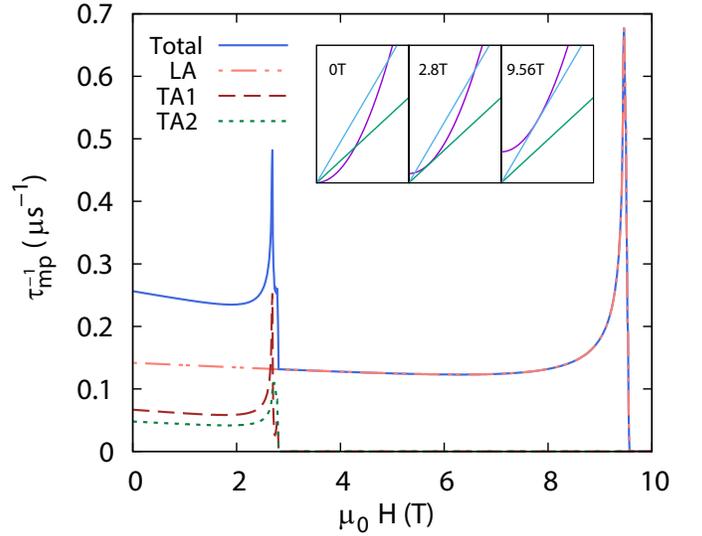}
\caption{(Color online) The magnon-phonon relaxation rate as a function of external magnetic field at high temperature limit. The inset shows the dispersion relation of  acoustic phonons and magnon as an illustration, where $\theta_k=\pi/2$.}
\label{fig_field}
\end{figure}

\begin{figure}[htb]
\includegraphics[width=\linewidth]{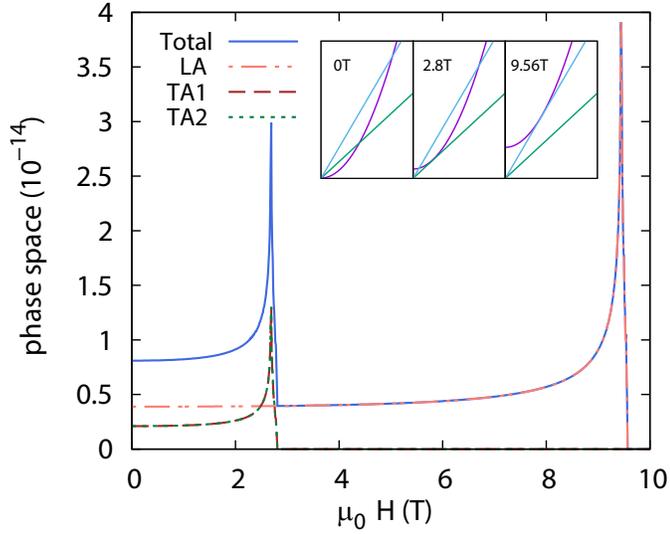}
\caption{(Color online) The phase space as a function of external magnetic field. The inset shows the dispersion relation of  acoustic phonons and magnon as an illustration, where $\theta_k=\pi/2$.}
\label{fig2}
\end{figure}
\begin{table}
\caption{Parameters of magnetoelastic coupling in YIG \cite{Flebus2017}}
\begin{tabular}{c c c c}
\hline
\hline
 & Symbol & Value & Unit \\
 \hline
 Lattice constant  & $a_0$ & 12.376 & $\text{\AA}$ \\
 Average mass density & $\bar{\rho}$ & 5.17 $\times$ 10$^{3}$ & Kg m$^{-3}$ \\ 
 Gyromagnetic constant & $\gamma$ & 2$\pi$ $\times$ 28 & GHz T$^{-1}$ \\
 Saturation magnetization & $\mu_0M_s$ & 0.2439
 & T \\
 Exchange stiffness & $D$ & 7.7$\times$ 10$^{-6}$ & m$^{2}$ s$^{-1}$ \\
 Magnetoelastic coupling & $B_{\perp}$ & 2$\pi$ $\times$ 1988 & GHz \\
 TA phonon velocity & $c_{1,2}$ & 3.9$\times$10$^3$ & m s$^{-1}$ \\
 LA phonon velocity & $c_3$ & 7.2$\times$10$^{3}$ & m s$^{-1}$ \\
\hline
\end{tabular}
\label{table1}
\end{table}

We present a numerical calculation of the relaxation rate
$\tau_{\text{mp}}^{-1}$ due to magnetoelastic coupling in YIG. The
parameters used in our calculation are listed in
Table.\,\ref{table1}. Fig.\,\ref{fig_temp} shows the temperature
dependence of relaxation rate in absence of magnetic
field. $\tau_{\text{mp}}^{-1}$ decreases rapidly with the increasing of
temperature and saturates above 100K, which is mainly attributed to
the temperature dependence of specific heat. In the high-temperature
limit ($\hbar\omega\ll k_{B}T$), $z^2e^z/(e^z-1)^2\rightarrow 1$,
$C_{\text{m}}\rightarrow Nk_{B}$ and $C_{\text{p}}\rightarrow
3Nk_{B}$, Eq.\,(\ref{Equ16}) could be reduced to: 
\begin{subequations}
\begin{eqnarray}
g_{\text{mp}}(T)=\frac{2\pi}{N\hbar^2}\sum\limits_{\kk\lambda} |M_{\kk, \kk \lambda}|^2\delta\left(\omega_{\text{m}}-\omega_{\text{p}}\right),
\end{eqnarray}
\begin{eqnarray}
g_{\text{pm}}(T)=\frac{2\pi}{3N\hbar^2}\sum\limits_{\kk\lambda} |M_{\kk, \kk \lambda}|^2\delta\left(\omega_{\text{m}}-\omega_{\text{p}}\right),
\end{eqnarray}
\end{subequations}
and it is obtained that $\tau_{\text{mp}}\rightarrow$ 0.26 $\mu
s^{-1}$ at high temperature limit. 

In the presence of magnetic field, the relaxation rate at high temperature limit are plotted
in Fig.\,\ref{fig_field} as a function of the strength of
magnetic field.  Magnetic field could effectively shift the
dispersion relation to higher energy. Thus
the intersects of magnon and phonon dispersions which satisfy the
energy conservation varies with the magnetic field. As a result, for
individual phonon modes, the relaxation rate is decreased with
the increasing of magnetic field, until reaching the critical magnetic
field $\mu_0 H=\frac{(\gamma D \mu_0 M_s
  \sin^2\theta-c_\lambda^2)^2}{4\gamma D c_{\lambda}^2}$. A sharp
increasing of relaxation rate is attributed to the tangency of phonon
dispersion and magnon dispersion, which maximize the interaction phase
space as shown in Fig.\,\ref{fig2}, where the phase space is defined
as $P_{mp}=\frac{1}{3N}\sum_{\kk\lambda}\delta(\omega_{\text{m}}(\kk)-\omega_{\text{p}}(\kk\lambda))$. It
is also noteworthy that one-magnon one-phonon interaction is forbidden
for TA phonon and LA phonon when $\mu_0 H>2.807$T and $\mu_0
H>9.567$T, respectively. 

In summary, we have theoretically studied the one-phonon one-magnon interaction in
ferrimagnet YIG. It has been obtained that one-phonon one-magnon
interaction is dominant at low temperature below $\sim 50$K or at the
resonant condition. It has also been verified that the two-temperature
model is valid for one-phonon one-magnon interaction. The
magnon-phonon relaxation is found to be tunable by the external
magnetic field. A maximum relaxation rate is obtained when the
tangency of phonon dispersion and magnon dispersion.

\acknowledgments
This work was supported by National Key R{\&}D
Program of China (No. 2017YFB0406004) and National Natural Science
Foundation of China (No. 11890703). 


\end{document}